\documentclass[journal,article,submit,pdftex,moreauthors]{Definitions/mdpi} 

\renewcommand{\linenumbers}{}

\usepackage{comment}

\firstpage{1} 
\pubvolume{1}
\issuenum{1}
\articlenumber{0}
\pubyear{2024}
\copyrightyear{2024}
\datereceived{ } 
\daterevised{ } % Comment out if no revised date
\dateaccepted{ } 
\datepublished{ } 
\hreflink{https://doi.org/} % If needed use \linebreak
\Title{TIPAA-SSL: Text Independent Phone-to-Audio Alignment based on Self-Supervised Learning and Knowledge Transfer}

\TitleCitation{TIPAA-SSL: Text Independent Phone-to-Audio Alignment based on Self-Supervised Learning and Knowledge Transfer}

\Author{Noé Tits $^{1}$, Prernna Bhatnagar $^{2}$ and Thierry Dutoit $^{3}$}

\AuthorNames{Noé Tits, Prernna Bhatnagar and Thierry Dutoit}

\AuthorCitation{ Tits, N.; Bhatnagar, P.; Dutoit, T.}
\address{%
$^{1}$ \quad University of Mons; noe.tits@umons.ac.be\\
$^{2}$ \quad University of Mons; prernna.bhatnagar@student.umons.ac.be\\
$^{3}$ \quad University of Mons; thierry.dutoit@umons.ac.be}

\abstract{In this paper, we present a novel approach for text independent phone-to-audio alignment based on phoneme recognition, representation learning and knowledge transfer. Our method leverages a self-supervised model (wav2vec2) fine-tuned for phoneme recognition using a Connectionist Temporal Classification (CTC) loss, a dimension reduction model and a frame-level phoneme classifier trained thanks to forced-alignment labels (using Montreal Forced Aligner) to produce multi-lingual phonetic representations, thus requiring minimal additional training. We evaluate our model using synthetic native data from the TIMIT dataset and the SCRIBE dataset for American and British English, respectively. Our proposed model outperforms the state-of-the-art (charsiu) in statistical metrics and has applications in language learning and speech processing systems. We leave experiments on other languages for future work but the design of the system makes it easily adaptable to other languages.}


\begin{document}

%%%%%%%%%%%%%%%%%%%%%%%%%%%%%%%%%%%%%%%%%%

\section{Introduction}
%%% Rephrase with Zoe-- Done 
Pronunciation is pivotal for language acquisition and effective communication. However, it often receives insufficient attention in second language (L2) education, leading to persistent challenges for learners in achieving clarity. Technology, particularly Computer-Assisted-Language-Learning (CALL) has emerged as a significant aid in supporting L2 pronunciation development across formal and informal learning environments~\cite{chapelle2009relationship}. Recent advancements in deep learning offer promising avenues for enhancing language learning by providing unlimited targeted and immediate feedback, addressing the class time and size limitations preventing educators from differentiating their pedagogy on such an individual skill. 

To improve pronunciation, integrating speech processing technologies with traditional teaching methods can provide a comprehensive approach to language learning. One fundamental task in this domain is text independent phone-to-audio alignment, a process that involves aligning phonetic representations with corresponding audio signals without relying on pre-determined text. This task is essential for accurately mapping the phonemes to their acoustic representations, contributing to the development of precise and effective speech processing technologies in language learning~\cite{golonka2014technologies, tits23_slate}.
%%%% Remarks from Thierry: again, a bit fuzzy. what do you mean exactly? I thought you meant "without text input" but you consider both cases next
Text independent phone-to-audio alignment faces a significant challenge due to the difficulty in obtaining extensive and well-annotated datasets. A possible solution to this challenge entails using established systems (like text dependent phone-to-audio alignment systems) to extract temporal information from publicly available speech datasets. This approach can be refined through the application of transfer learning and self-supervised learning methodologies in the development of a solution.

%% Rewrite all of this
Transfer learning~\cite{tan2018survey_deep_transfer_learning}, is a method within the field of deep learning. Its approach involves pre-training a model on a large dataset and subsequently fine-tuning it on a smaller dataset tailored to the specific task. This methodology has demonstrated considerable success across various domains, particularly in the context of self-supervised learning.

In Self-supervised learning~\cite{jaiswal2020survey} a model is trained to autonomously learn representations of input data without the need for explicit supervision. This proves particularly advantageous when labeled data is either limited or entirely unavailable. Within the field of speech technology, the application of self-supervised learning through transfer learning has proven invaluable in addressing several complex scenarios. For example, in Automatic Speech Recognition (ASR) for low resource languages~\cite{zoph2016transfer}, where annotated data may be scarce, transfer learning allows the utilization of pre-trained models on more data-abundant languages, adapting them effectively to the target language. Likewise, in emotional or expressive speech synthesis~\cite{exploring_transfer_learning-19-tits, visualization-19-tits, tits2021analysis}, speech emotion recognition~\cite{asr-based-features-18-tits}, voice conversion~\cite{zhou2022emotional} and pronunciation assessment~\cite{hu2015improved}.

In this paper, we take full advantage of state-of-the-art methodologies in deep learning, self-supervised learning, and phonetic representation to present a novel approach to text independent phone-to-audio alignment. 
%%Comments from Thierry: Do not use the word "leveraging" so often --- Done
%% Why are we doing this?? 
%% SOTA: Charsiu : only US English 
%% Possible solution:: Use other varities of English. Pedagogical needs. So reframe for that. 
Most state-of-the-art self-supervised systems perform well on American English which means that other varieties of English get penalised and the model is biased towards American English. The rationale to develop the system proposed in the paper was the pedagogical needs to create a system that performs equally well on other variants of English as it does on American English. For our system, we use the self-supervised model Wav2Vec2, fine-tuned for phoneme recognition using CTC loss. We integrate this with a dimensional reduction model based on Principal Component Analysis (PCA) and a frame-level phoneme classifier. The resulting model pipeline produces a vector of probabilities for every audio frame. From the same model we also extract predicted phonemes and their boundaries and hence use this information for a text independent phone alignment system. 

%Continue here 

%In addition to exploring the landscape of phoneme recognition and phone-to-audio alignment, we introduce a comparative analysis that includes traditional HMM-based systems, text-dependent alignment tools, and innovative approaches predicting both phones and their temporal boundaries. Through extensive evaluation using synthetic native data, encompassing diverse voices and accents, including US/UK Female/Male voices, we demonstrate the robustness and language independence of our proposed method.

%Our contribution lies in providing a unified framework that combines the strengths of self-supervised learning, phoneme recognition, and representation learning for text-independent phone-to-audio alignment. This not only advances the state-of-the-art in this specific task but also opens avenues for broader applications in language learning and speech processing systems. The following sections detail the related work, methodology, experiments and results; we then conclude about the findings of our approach. 

The major contributions of this paper are as follows: First,
we propose a text independent phoneme alignment system using self-supervised learning. This not only advances the state-of-the-art in this specific task but also opens avenues for broader applications in language learning and speech processing systems. Second, this system functions effectively with diverse English language variations (British) and is capable of accommodating various languages, making it language-independent.

\section{Related Work}
\subsection{Phone-to-Audio Alignment}

Text independent phone-to-audio alignment involves predicting a sequence of phones and their temporal locations within speech signal without prior linguistic information, such as a known text or phone sequence.

In contrast, text dependent phone-to-audio alignment utilizes text information to align a phone sequence generated from a grapheme-to-phoneme model applied to textual inputs.\\
%%%%%% Insert here information provided by Sandrine: As for HMM alignment, there's really no reason it shouldn't work out without a prior transcription. Since HMMs do good recognition, you just need to do recognition and you will get the alignment. .%%%%
Hidden Markov Models (HMMs) have traditionally played a prominent role in aligning  phonetic and temporal information, including but not limited to Kaldi~\cite{povey2011kaldi} and HTK~\cite{young2002htk}. Forced alignment systems using HMMs include Gentle\footnote{\url{https://lowerquality.com/gentle/}} and ProsodyLab~\cite{gorman2011prosodylab}. However, such models face limitations when attempting to predict both phones and phone boundaries simultaneously. The challenges for HMMs become apparent when confronted with long-range dependencies within a sequence. However, they struggle to effectively capture and understand the broader context of a word within a sentence. As a result, this hinders their ability to grasp nuanced linguistic relationships and context. Another reason why HMMs may not perform well is due to the incorrect phonetic transcriptions due to variations in conversational speech.\\

Recent developments have witnessed a shift towards deep learning models like Recurrent Neural Networks (RNNs)~\cite{koizumi1996recurrent} and models trained with CTC loss~\cite{muller2017phonemic}, for phoneme recognition. This shift towards deep learning models allows us to efficiently predict both phones and phone boundaries simultaneously. \\

\subsection{Phoneme Recognition}
Deep learning models are known for their ability to capture complex patterns in data and are frequently employed for phoneme recognition. CTC loss is a popular training criterion for such models in this context. It allows the model to learn the alignment between the input speech signal and the corresponding phoneme sequence, even when the temporal correspondence is not provided during training. Wav2Vec2~\cite{conneau2020unsupervised, xu2021simple} stands out as a self-supervised model that has been fine-tuned specifically for phoneme recognition using the CTC loss.

One of the advantages of the Wav2Vec2 approach is its ability to generate multi-lingual phonetic representations. By leveraging self-supervised learning during pre-training, the model learns to extract robust features from speech signals, capturing phonetic information that is broadly applicable across different languages. Furthermore, the fine-tuning process with CTC loss refines the model's ability to map these learned representations to specific phoneme sequences.\\

\subsection{Systems Predicting Phones and Boundaries}

%charsiu~\cite{charsiu-zhu2022phone} stands out as the only work providing a system capable of predicting both phones and phone boundaries. This integrated approach contributes to a more comprehensive understanding of the speech signal.

%%%%% Comments from Thierry: impossible. Most HMM based systems do both. I don't see why you  sasy s this, or don't understand what you mean then.
%%%%%Rephrased
%%% PB: Not changig this till we get some confirmation from Thierry 
Charsiu~\cite{charsiu-zhu2022phone} is recognized for its unique capability to predict both phones and phone boundaries. This integrated approach contributes to a more comprehensive understanding of the speech signal. While HMMs can perform similar tasks, they are often employed more prevalently for forced-alignment tasks, particularly in text-dependent scenarios. The emphasis on forced alignment implies that these systems are commonly utilized when the corresponding phoneme or phone sequence is available.

Therefore, in this paper, we conduct a comparative analysis between our proposed model and the text independent aspect of the charsiu model, recognized as the state-of-the-art for this task. Our methodology, detailed in the following section, combines self-supervised learning with phoneme recognition.

\section{Methodology}
%%Comments from Thierry: mention Figure 1 here. Also "the system" shoudl at least be "our system" notice the title "system" is fuzzy too; be more precise.
\subsection{Our Proposed Method}
Our proposed method is an innovative approach, using wav2vec2, Principal Component Analysis (PCA) for dimensional reduction and frame-level phoneme classification, offers a robust text independent phone-to-audio alignment. The system is explained in details in subsection 3.3 and the system's architecture is depicted in Figure 1. The model's robustness is demonstrated through evaluation using synthetic native data, using the TIMIT~\cite{garofolo1993timit} and SCRIBE dataset. Thanks to its versatility, we expect that our method will find applications in language learning and speech processing systems.

%Section~\ref{sec:training} details the training stages of the system, then Section~\ref{sec:inference} described the different processing steps of the pipeline at inference.

\begin{figure}[h!]
  \centering
  \includegraphics[width=\linewidth]{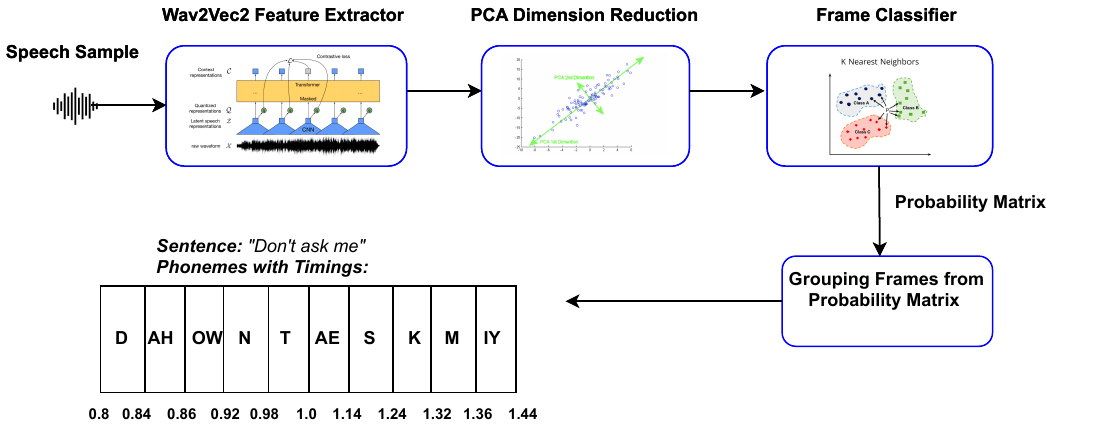}
  \caption{Model pipeline of the system}
  \label{fig:tech_pipeline}
\end{figure}
%%% Edit the image :: so you need text or phonetic input??

%\subsection{Training}
%\label{sec:training}
%The training of our system consists of two stages: 
%%%%% Comments from Thierry: i don't understand your 2 stages
%\begin{itemize}
   % \item Data extraction: speech representations, and phone timings
    %\item Model pipeline training through knowledge transfer to obtain a frame-level phoneme classification
%\end{itemize}

\subsection{Pre-trained Self-Supervised Model}
\label{sec:data_extraction}
%\subsubsection{Wav2vec2 representations}

In this Section, we explain what open-source model we use as a basis\footnote{\url{https://huggingface.co/facebook/wav2vec2-xlsr-53-espeak-cv-ft}} in order to leverage learned cross-lingual representations of speech adapted towards the phonetic space.

%%% TO DO: Define quantized codebook, what is c and q 
The basis of this model is a self-supervised model (wav2vec2) trained to extract latent representations from the audio data. The wav2vec2 model is trained on a large dataset of unlabeled speech data, using a contrastive predictive coding loss function to learn a representation that is capable of predicting future audio frames and can then be utilized for downstream tasks. The pre-training loss is defined as \cite{baevski2020wav2vec}:

\begin{equation*}
    L= L_m+ \alpha L_d
\end{equation*}
Where, $L_m$ is contrastive loss, $L_d$ is diversity loss and $\alpha$ is the hypertuned parameter
\begin{equation}
{L}_m=-\log \frac{\exp \left(\operatorname{sim}\left(\mathbf{c}_t, \mathbf{q}_t\right) / \kappa\right)}{\sum_{\tilde{\mathbf{q}} \sim \mathbf{Q}_t} \exp \left(\operatorname{sim}\left(\mathbf{c}_t, \tilde{\mathbf{q}}\right) / \kappa\right)}
\end{equation}

% Rephrased: 
In the above equation, $k$ represents the fixed temperature, and sim signifies the cosine similarity between context representations and quantized latent speech representations. The term $L_m$ resembles the Softmax Function, but it employs cosine similarity instead of a score. For ease of optimization, the negative logarithm of the fraction is also applied. $\textbf{c}_t$ is the context network output centered over masked time step $t$ and $\textbf{q}_t$ is the true quantized latent speech representation.

In this paper we use the version of the model pre-trained on cross-lingual speech data (53 different languages) \cite{conneau2020unsupervised} that was then  fine-tuned \cite{xu2021simple} for the task of phoneme sequence prediction using a CTC loss, and the result is an open-source \footnote{\url{https://huggingface.co/facebook/wav2vec2-xlsr-53-espeak-cv-ft}}.
%% Define CTC Loss using an equation. ---- Done

The model itself thus predict sequences of phoneme labels, but no phone boundaries. What we propose in the following sections is an approach for leveraging the learned cross-lingual speech representations of this model that was already oriented to a phonetic space thanks to the CTC fine-tuning. The goal of this approach is to require limited amount of data and to be less biased towards American English accent compared to existing models.

%\subsubsection{Data balancing}
\subsection{Data Processing and Knowledge transfer}

%%%^^^ Maybe remove this figure, if not, mention which dataset and which language 
%%%% ---- Done 
Based on the representations that we can extract from the model explained in the previous section, we used PCA to reduce dimensionality while retaining most of the variance in the data. We use the last hidden layer before classification heads as our speech representations.

We aimed to preserve a substantial amount of variance of the data. Experimenting without PCA and with varying levels of variance retention (99\%, 95\%, and 90\%), we determined that retaining 95\% offers the most favorable compromise in terms of classification results. We hypothesized that this choice may serve as a means of noise filtration and provided a more manageable space for classifiers to process effectively. To train this dimension reduction model and the subsequent frame classifier model we used the MAILABS~\cite{solak2019m} dataset
which is a large dataset of speech data, containing recordings from various languages and accents to train our reducer and classifier. We used the American and British English datasets from MAILABS.\\

%%%%% Reframe this sentence %%%%%%
We worked at the frame level for training the shallow dimension reduction and classifier models. As we observe from Figure 2, the frequency of phonemes in natural languages is not uniform. Hence, it is essential to perform data balancing on the data extracted from MAILABS for unbiased training. To perform data balancing, we randomly selected an equal number of latent frames in order to have a balanced set of labelled latent frames for every phoneme.\\
\begin{figure}[ht]
  \centering
  \includegraphics[width=\linewidth]{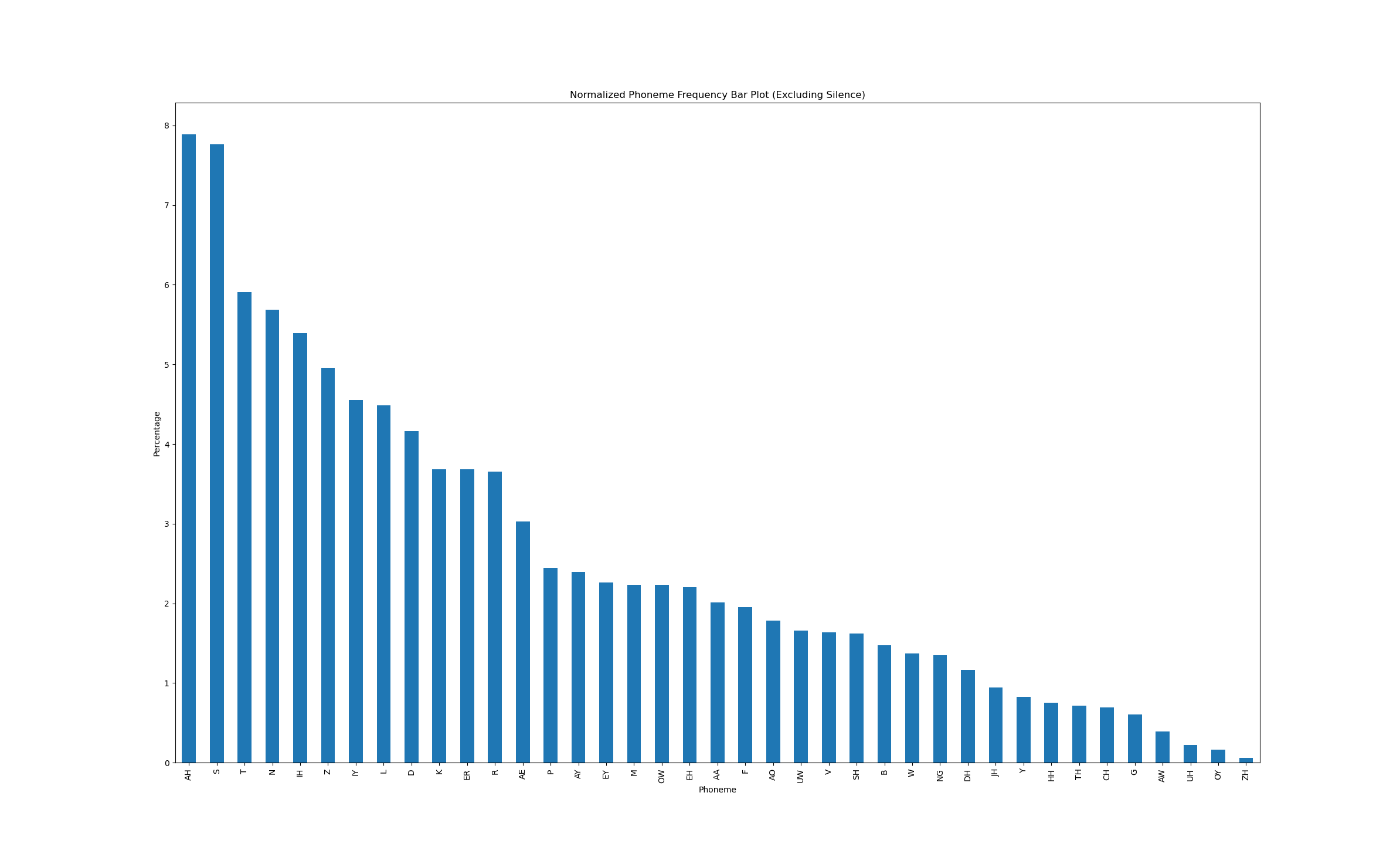}
  \caption{Frequency of Phonemes in MAILABS for English}
\end{figure}

%%% Add some more information about PCA here 
%%%Noise reduction :  PCA can help filter out noise or irrelevant information present in the data by focusing on the most significant components. This can lead to improved model performance, especially in scenarios where noisy features may adversely impact the results.
%%Data Pre  processing:: It is often used as a pre processing step before applying the other ML algorithms. By reducing the dimensionality of the data , PCA can improve the performance and efficiency of subsequent analyses and models
Finally, we trained a frame-level phoneme classifier based on this reduced space. To train this model, we thus built a reduced latent frame dataset from information extracted from MAILABS dataset. For the classifier we use K-Nearest Neighbors (KNN) with 10 neighbours. The frame classifier produced a probability matrix, with each row representing a frame and its corresponding predicted probability for a phoneme. We selected the phoneme which has the maximum posterior probability for each frame. Subsequently, consecutive frames with identical phonemes were grouped together and the start and end timings of each group was computed using frame indices and timestamps. To filter noise, we applied a threshold of 0.5 over the probabilities to remove phonemes that have less than the threshold value. After this step, we merged the consecutive groups to obtain the final timings of the audio sequence.\\
%%%% Thierry: OK but again be precise : how many PCAs, decribe what you mean by "train a PCA" better.---- Done 
Our proposed approach, which integrates a reduction model with a classifier model can adeptly model intricate relationships stemming from the speech representation learning model. The task of mapping between these learned representations and phonetic units is efficiently managed by a compact and effective model. This ensures our ability to comprehend the complex relationships within the data while maintaining computational efficiency.\\

\section{Experiments}

In this section, we describe the experiments conducted to evaluate the performance of our proposed phone-to-audio alignment model. 

\subsection{Text independent phone-to-audio alignment on TIMIT}
\begin{table}[ht]
\centering
\scriptsize
\caption{Evaluation results of text independent alignment on TIMIT dataset}
\begin{tabular}{lccccc}
\hline
Model & P & R & F1 & r-val \\ \hline
\texttt{W2V2-CTC-10ms}   & 0.31  & 0.29  & 0.30   &  0.42          \\ 
\texttt{W2V2-CTC-20ms}   & 0.31  & 0.30 & 0.31   &  0.42        \\ \hline
\multicolumn{6}{l}{\textit{Phone recognition} + \texttt{W2V2-FS}} \\\hline
\texttt{W2V2-FS-20ms} & 0.40 & 0.42 & 0.41 & 0.48  \\
\texttt{W2V2-FS-10ms} & 0.56 & 0.58 & 0.57 & 0.63  \\
\texttt{W2V2-FC-32k-Libris}  &  0.57 & 0.57 & 0.57 & 0.64  \\ \hline
\textit{Direct inference } \\\hline
\texttt{W2V2-FC-20ms-Libris}  & 0.57  &  0.59 & 0.58   &   0.63  &      \\
\texttt{W2V2-FC-10ms-Libris}  & 0.55  &  0.58 & 0.56   &   0.62  &      \\
\texttt{W2V2-FC-32k-Libris}  & \textbf{0.60}  & \textbf{0.63}  & \textbf{0.61}   &  \textbf{0.66}  &     \\ \hline
\textit{\textbf{Our Proposed Model}} \\\hline
\texttt{W2V2-PCA-C}  &\textbf{ 0.61}  & \textbf{0.68} & \textbf{0.63}   &   \textbf{0.58}    \\
\end{tabular}
\label{tab:inde}
\end{table}

In evaluating the performance of our text independent phone-to-audio alignment model, we conducted a comprehensive comparison with the state-of-the-art model, charsiu~\cite{charsiu-zhu2022phone}.
%% Explain charsiu a bit 
In the charsiu model the authors compare two systems: text independent phone-to-audio alignment and text dependent phone-to-audio alignment using Wav2Vec2. The Wav2Vec2-FS is a semi-supervised model which learns alignment using contrastive learning and a forward sum loss. The second model, Wav2Vec2-FC, is a frame classification model trained on labels for forced alignment, capable of both forced alignment and text-independent segmentation. The evaluation of both the systems for charsiu has been performed on the TIMIT dataset. It has been used for this evaluation because of the availability of human annotations especially phoneme level. We use a similar approach to evaluate the same metrics for our model. 
The assessment is based on statistical measures, namely precision, recall, F1 score, and r-value. In the referenced work, the authors present their best-performing model, W2V2-FC-32k-Libris, as can be observed in Table 1.
%%%% Thierry: why is this impressive? these are low numbers for a classifier. Remove the word "impressive" but this is SOTA. C'est la vie. ---- Done 
\\

%%%%% Thierry: you did not mention TIMIT for CHARSUI. you shoud :)--- Done 

% Data and codes for data processing
%https://www.kaggle.com/datasets/mfekadu/darpa-timit-acousticphonetic-continuous-speech/data

In our case, for the task of phoneme recognition, we compare our model with the text independent W2V2-FC-10ms charsiu model. The statistical metrics for our proposed model outperform that of the charsiu model. We compare the TIMIT test dataset to assess our model and we observe from Table 1 that the r-value for our model deteriorates in performance. Apart from r-value, all the other metrics: Precision, Recall and F1 value have shown to perform well for our model.\\

The tests for our model are also performed on the TIMIT dataset but has the possibility to extend to other datasets with real speech and also other languages. \\

The r-value, which is known as the Pearson correlation coefficient measures the similarity between the ground truth phonemes and the predicted phonemes. Some of the reasons for low r-value could be alignment errors, variability in the pronunciation of speakers or changes in speaking style within the dataset. However, we need more experiments to confirm this hypothesis.  
%%%%% TODO: 27.02.2024 
%%% 1. Explain why we use TIMIT --- Done. Maybe more?
%%% 2. Introduce Scribe and explain why we use it

\subsection{Text-independent phone-to-audio alignment on SCRIBE}
A second part of the experiments in our proposed model is to evaluate it on British English. 

The SCRIBE\footnote{\url{https://www.phon.ucl.ac.uk/resource/scribe/}} dataset is a small corpus of read speech and spontaneous speech specializing in British English. It consists of 200 'phonetically rich' sentences and 460 'phonetically compact' sentences. The 'phonetically rich' sentences are  phonetically balanced. The 'phonetically compact' sentences are based on a British version of the MIT compact sentences (as in TIMIT). There are 45 files in the dataset of 30-50s containing 5-10 sentences. 

The audio files and the phoneme annotations needed some processing before we could start using the dataset. The phonemes are in SAMPA form and needed to be converted to the English Arpabet. Even after the conversion from SAMPA to Arpabet, there were symbols that we could not retrieve so we filtered them out. 

We evaluated the charsiu  model and our proposed model using SCRIBE and achieved the metrics that are depicted in Table 2. 
\begin{table}[h]
\centering
\scriptsize
\caption{Evaluation results of text-independent alignment on SCRIBE dataset}
\begin{tabular}{lccccc}
\hline

Model & P & R & F1 & r-val \\ \hline

\textbf{charsiu Model} \\\hline
%\texttt{W2V2-FC-20ms-Libris}  & 0.57  &  0.59 & 0.58   &   0.63  &      \\
\texttt{W2V2-FC-10ms-Libris}  & 0.93  &  0.71 & 0.80   &   0.79  &  
%\texttt{W2V2-FC-32k-Libris}  & \textbf{0.60}  & \textbf{0.63}  & \textbf{0.61}   &  \textbf{0.66}  &     
\\ \hline
\textbf{Our Proposed Model} \\\hline
\texttt{W2V2-PCA-C}  &\textbf{ 0.89}  & \textbf{0.85} & \textbf{0.87}   &   \textbf{0.88}    \\
\end{tabular}
\end{table}

Upon close examination of the results presented in Table 2, it becomes evident that the metrics associated with our proposed model exhibit a greater uniformity in values when compared to charsiu. Furthermore, these metrics generally surpass those of charsiu, albeit with the exception of precision. The uniformity observed in our model's metrics can be attributed to the utilization of a more balanced reduced frame dataset during training, which serves to mitigate biases and yield more consistent outcomes. Moreover, the superior quality of audio files within the SCRIBE dataset, resembling professional studio recordings with minimal noise, likely contributes to the heightened metrics observed for both charsiu and our proposed model when contrasted with the TIMIT dataset.

Consequently, our system demonstrates enhanced generalization across various accents, laying the foundation for potential expansion to other languages. The key lies in leveraging the underlying Wav2vec2 XLSR framework, and the methodology employed with MAILABS can seamlessly be replicated with different datasets, opening avenues for utilizing non-native English accents and broader linguistic applications.

\section{Discussion}

%% So, the conclusion should consist of ::
%Limitations
%Future Work
%Conclusions
%

In this paper, we introduced an innovative text independent phone alignment system designed to be language-independent. Our method harnesses a self-supervised model (wav2vec2) fine-tuned for phoneme recognition through a CTC loss, alongside a dimensional reduction model (PCA) and a frame-level phoneme classifier. Through experiments using synthetic native data, we assessed our model's performance on both American and British English, benchmarking it against the state-of-the-art charsiu model. Encouragingly, our results demonstrate robustness and applicability to diverse accents, such as British English.

However, certain limitations are acknowledged. Firstly, the reducer and classifier have been trained on a restricted amount of data specific to native English, posing a constraint. Secondly, our model relies on forced-aligned datasets of native speech for effective learning, introducing another limitation. Thirdly, the SCRIBE dataset that we used is a smaller dataset of British English as compared to the TIMTI dataset. Lack of well-annotated datasets, especially on phoneme level is one of our biggest challenges.

Future research directions could explore the incorporation of datasets containing non-native English data. This involves re-training the shallow reducer and classifier models using non-native speech data. Furthermore, extending our approach to languages beyond English and evaluating its performance with real-world data from language learners presents an intriguing avenue for exploration. Our work lays the foundation for further experiments in the realm of text independent phone-to-audio alignment, especially in the context of non-native English.

\funding{
Part of this work was done during the project \textit{REDCALL} that is partially funded by a FIRST Entreprise Docteur program from SPW Recherche\footnote{https://recherche.wallonie.be/}.

Part of this work was done during the project \textit{DEEPCALL} that is partially funded by a Win4Doc program from SPW Recherche.
%Please add: ``This research received no external funding'' or ``This research was funded by NAME OF FUNDER grant number XXX.'' and  and ``The APC was funded by XXX''. Check carefully that the details given are accurate and use the standard spelling of funding agency names at \url{https://search.crossref.org/funding}, any errors may affect your future funding.
}

%%%%%%%%%%%%%%%%%%%%%%%%%%%%%%%%%%%%%%%%%%
\vspace{6pt} 

\begin{adjustwidth}{-\extralength}{0cm}

\reftitle{References}


\bibliography{biblio}

\end{adjustwidth}

\end{document}